
\magnification=1200
\settabs 18 \columns

\baselineskip=17 pt
\topinsert \vskip 1.00 in
\endinsert

\font\steptwo=cmr10 scaled\magstep2

\def\b{\bigskip}
\def\bb{\bigskip\bigskip}

\def\sqr#1#2{{\vcenter{\vbox{\hrule height.#2pt
 \hbox{\vrule width.#2pt height#1pt \kern#1pt
 \vrule width.#2pt} \hrule height.#2pt}}}}

\def\operp{\hbox{${\kern+.25em{\bigcirc}
\kern-.85em\bot\kern+.85em\kern-.25em}$}}

\def\lsim{\;\raise0.3ex\hbox{$<$\kern-0.75em\raise-1.1ex\hbox{$\sim$}}\;}
\def\gsim{\;\raise0.3ex\hbox{$>$\kern-0.75em\raise-1.1ex\hbox{$\sim$}}\;}
\def\no{\noindent}
\def\r{\rightline}
\def\ce{\centerline}
\def\ve{\vfill\eject}
\def\rdots{\mathinner{\mkern1mu\raise1pt\vbox{\kern7pt\hbox{.}}\mkern2mu
 \raise4pt\hbox{.}\mkern2mu\raise7pt\hbox{.}\mkern1mu}}

\def\e e{$e^+ e^-$ }

\def\to{\rightarrow}
 \voffset=-1truein
\r {hep-th/9712239}
\r {CERN-TH/97-375}
\r {UCLA/97/TEP/32}
\def\today{\ifcase\month\or January\or February\or March\or April\or
May\or June\or July\or August\or September\or October\or November\or
December\fi \space\number\day, \number\year }

\r \today
\bb\bb\bb

\vskip.5truein

 \ce{\bf CONFORMAL MAXWELL THEORY AS A SINGLETON FIELD THEORY}
\ce{\bf ON ADS$_5\,$, IIB THREE BRANES AND DUALITY}
\vskip.5cm

\ce{Sergio Ferrara* and Christian Fronsdal**}
\ce{\it *CERN, Geneva, Switzerland.}
\ce{\it **Physics Department, University of California, Los Angeles, CA
90090-1547.}
\vskip1.0cm

\baselineskip=12pt
\no ABSTRACT.  We examine the boundary conditions associated with extended
supersymmetric Maxwell theory in 5-dimensional anti-De Sitter space.
Excitations on
the boundary are identical to those of ordinary 4-dimensional conformal
invariant super electrodynammics.  Extrapolations of these excitations give
rise to a 5-dimensional topological gauge theory of the singleton type.
The possibility of a connection of this phenomenon to the world volume
theory of 3-branes
in IIB string theory is discussed.

\ve

\baselineskip=17pt

\voffset=0truein

\line{{\bf I. Introduction} \hfil}

Recent developments in the theory of $p$-branes, and duality interconnections
in $M$ theory and string theories, have brought renewed interest in higher
dimensional supergravity theories and their extensions [1][2].
In recent times attention has been called to an  intriguing connection,
between the horizon geometry of certain black $p$-brane configurations and
the world volume dynamics of the brane degrees of freedom [3].
Further relations between Anti-de Sitter supergravities and brane backgrounds,
using S- and T-duality, have been also investigated [42].

This connection follows from earlier considerations [4][5][6], where it was
pointed out that the asymptotic horizon
geometry of certain $p$-branes in $d$ dimensions, typically on an
ADS$_{p+2}\times S_{d-p-2}$ background, admits a superalgebra that is
identical to the superconformal algebra of the corresponding world-volume
$p+1$ dimensional field theory, the latter describing the world
volume degrees of freedom when gravity (or substringy effects) is
decoupled. \footnote*{The full nonlinear Born-Infeld type action is presumably
conformally invariant if coupled to the appropriate anti-De Sitter
background [3][10].}
Moreover, the horizon geometry of the $p$-branes has twice the number of
supersymmetries carried by the
brane background far from the horizon [6][7][8]. In the cases discussed
below the relevant superalgebras admit
32  supersymmetries.

The most prominent examples   are the $M$-theory (11 dimensional
supergravity [9]) chiral
five-brane, [10] where the underlying superconformal symmetry is
$Osp(6,2/4)$ [11], with
bosonic isometry $O(6,2)\times USp(4)$, and the Dirichlet three-brane of
the IIB
superstring, where the underlying superconformal symmetry is $SU(2,2/4)$
[11][12][13][14][15], with
bosonic isometry $SU(2,2)\times SU(4) \approx O(4,2)\times O(6)$.

These conformal symmetries are the same as the isometries of the
corresponding asymptotic horizon background, i.e. $AdS_7\times S_4$ and
$AdS_5\times S_5$ respectively.  This mathematical coincidence was the
origin of a proposed duality [3] between supergravity around the horizon
background and superconformal brane dynamics.  It is a kind of strong-weak
coupling duality in which the fundamental supergravity degrees of freedom
(supergravitons) are expected to emerge as bound states in the
non-perturbation regime of the corresponding (world-volume) theory.

It is the aim of the present work to further investigate this
idea, also reminiscent of previous speculations [5][6][16][17][18] on the
interpretation of the conformal world-volume degrees of freedom as ``boundary"
degrees of freedom of certain field theories formulated in anti-De Sitter
space.  The latter have the unconventional property that they admit
``topological" unitary representations [19][20][21][38], the singletons (or
their
ramifications) that do not correspond to local degrees of freedom  in the
bulk, but to interactions purely localized on the $p+1$-dimensional boundary
of AdS$_{p+2}$.

In this context it was also
proposed [4][5][16][18] that 2-brane dynamics (the fundamental brane of
$M$-theory) should correspond to
singleton field theories in AdS$_4\times S_7$, with underlying superalgebra
$OSp(8/4)$.
The corresponding supergravity is gauged $SO(8)$ in $d = 4$ [22].
Similarly, 1-branes corresponding to type IIA, type IIB and heterotic
fundamental strings were interpreted [17] as singleton field theories of
the 3-dimensional conformal
group $SO(2,2) \approx Sp(2,{\bf R}) \otimes Sp(2,{\bf R})$ with
superconformal algebras
given respectively by $Osp(8/2)_c \otimes Osp(8/2)_s, Osp(8/2)_c \otimes
Osp(8/2)_c$ and
$Osp(8/2)_c\otimes Sp(2,{\bf R})$.

In the present situation  the corresponding supersingleton degrees of freedom
are the (2,0), $d=6$ tensor multiplet and the $N=4$, $d=4$ vector multiplet
($U(N)$ Yang-Mills for $N$ three branes).
\footnote *{More precisely, the singleton field theory corresponds to the
Goldstone multiplet on the branes [6][10]. For a $U(N)$ theory, this is the
$U(1)$ Maxwell multiplet
of the $N = 4, d = 4$ superconformal algebra, describing the center of mass
motion of the 3-brane.}
It was
already shown in [21], that compactification of 11 dimensional supergravity on
$S_4$ down to seven dimensions [23],  gives rise to a spectrum of unitary
representations containing a
singleton representation, called doubleton by in [21], which precisely
correspond to a self-dual 2-tensor, a Weyl spinor and 5 scalar degrees of
freedom.  It is crucial to the present work that  these degrees of
freedom are purely topological in the bulk, they do not allow local interaction
in AdS$_7$ but only on its six-dimensional boundary [10], where they are
conformal.

The same analysis was carried out for the case of $S_5$ compactification of
IIB, 10-dimensional supergravity on five-dimensional anti-de Sitter space
[38].  This
is the supergravity theory that corresponds to the horizon geometry of the IIB
three-brane [3].  The corresponding sueprgravity theory is gauged 5-dimensional
supergravity with $SU(4)$ gauge symmetry [25].

In this case the super-singleton multiplet (called $SU(2,2/4)$ doubleton
in[38]), corresponds to the $N=4$ super-Yang-Mills multiplet.

In the present paper we will extend this analysis,
showing in particular that the $N=4$ vector multiplet does indeed have the
property that its
degrees of freedom are purely topological, in the sense
that they do not lead to local interaction on $AdS_5$, but only to
interactions on the 4-dimensional boundary where
the action of $O(4,2)$ becomes the conformal group of 4-dimensional field
theories [24].

Our analysis will essentially consist of a proof that conformal
invariant Maxwell theory can be interpreted as a topological ``singleton" field
theory in $AdS_5$. It is done by Fourier analysis of the propagator, as in
earlier studies of 4-dimensional singletons [19].

The paper is organized as follows:  In Section II we recall the theory that is
dual, in the sense of  [3], to the world volume 3-brane theory in the
conformal limit.  This is a gauged, $d=5$ maximal supergravity [25] with 32
supersymmetries.  This theory can be defined {\it per se} or can be viewed
as coming from the IIB string, when compactified on a five-sphere
down to five dimensions [25][3].  In Section III we give preliminaries of field
equations in anti-De Sitter space times.  In Section IV we formulate
conformal Maxwell theory and its relation to a singleton field theory in
AdS$_5$.  In Section V we consider superconformal   and
$N=4$ extended superconformal Maxwell theory and their interpretation
 as singleton field
theories in AdS$_5$.  The paper ends with a paragraph of conclusions and
suggestions for future investigations.
\bb

\line{{\bf II. Gauged Supergravity in anti-De Sitter Space.} \hfil}

Gauged $N=8$ supergravity in five dimensions was constructed by Gunaydin,
Romans and Warner in 1984 [25].  It lives in anti-De Sitter space-time
$O(4,2)/O(4,1)$ and admits 32 supersymmetries.  It can be viewed as the
gauge theory of the superalgebra $SU(2,2/4)$, the same superalgebra as that
of its
``dual", namely $N=4, d = 4$ super-Yang-Mills theory.

We assume, following [3], that the supergraviton multiplet arises as a
bound state in the dual superconformal theory.  Such
bound states form a multiplet, which corresponds to a unitary massless
representation of $SU(2,2/4)$ containing 1 graviton, 8 gravitini, 15 gauge
bosons, 12 antisymmetric tensors, 48 spin-${1\over 2}$ fermions and 42
scalars.  Note
that there is a relation [25][26] between the gauge group $SU(4)$ and the
5-dimensional duality
group $E_{6(6)}$ of toroidal compactification [27].
Indeed $SU(4)$ is the subgroup of the maximal compact  subgroup $USp(8)$ in
$E_{6(6)}$, on which the reduction of the 27-dimensional representation of
$USp(8)$
contains the adjoint representation.  Indeed $27\to 15+6\times 2$ as
$USp(8)\to SU(4)$ (this corresponds to the embedding $ 8 \rightarrow 4 +
\bar 4$).  Note that the theory has
a potential [25] for the 42 scalars, which decomposes as follows under $SU(4)$
 [43]:
$42 \to 20 + 10 + \bar{10} + 1 + \bar{1}$.
This is consistent with the fact that 20 and 10 are ``massive', since their v.e.v. is driven to zero by $SU(4)$ symmetry, while the singlet is massless
since its quadratic fluctuation is absent from the potential.
(This analysis is similar to gauged $N=8$, $d=4$ supergravity in the $SO(8)$
invariant vacuum.)

It is most important that the supergravity theory in AdS$_5$  also has a
global $SU(1,1)$ symmetry, to be identified with the $S$-duality in the
corresponding dual super Yang-Mills theory.  Note that here $SU(1,1)$ is
the subgroup of $E_{6(6)}$ that commutes with the gauge group
$SU(4)$.  This symmetry is the original $SU(1,1)$ symmetry of  type IIB,
10-dimensional
supergravity [28].  Note the chain decomposition [25]:
$$
E_{6(6)} \to SL(6,R)\times SL(2,R)\to SO(6)\times SL(2,R)~.
$$
\no (Of course, the two subgroups $SL(2,R)$  and $USp(8)$ do not commute.)
It is easy to identify the
``composite operators" of the world-volume theory that correspond to the
supergraviton in $AdS_5\times S_5$.  They are contained in the supercurrent
[29], $N=4$ multiplet that
contains two $N=4$, spin ${\scriptstyle {3\over 2}}$ currents, giving 8
spin-3/2 gravitinos, the stress tensor giving the
graviton, and 15
$SU(4)$ ($R$-symmetry) currents giving the anti-De Sitter gauge bosons.
It also contains the scalars in the 20 of $SU(4)$ as its first components.

We also note that in the anti-De Sitter supergravity context, according to
[25], one gets a
quantization of the ratio $k'/g^3$ where $k'$ is the 5-dimensional gravitional
constant and $g$ is the 5-dimensional gauge coupling.  This is due to
Chern-Simons
interactions\footnote *{ Turning things around, 5-dimensional gauged supergravity may
be viewed as the supersymmetrization of the
5-dimensional Chern-Simons coupling. A similar argument applies to gauged
7-dimensional supergravity [23]. This
theory lives in AdS$_7$ and has gauge group $USp(4)$. It is ``dual" to the
$M$-theory five-brane [3][10].
 \vskip-2mm} and to the non-triviality of the 5'th homotopy group of $SU(4)$.
\footnote{**}{  Strictly speaking, the
quantization would only apply if the 5-manifold were compact, while AdS$_5$
has the topology of $S_3 \times S_1 \times
R^+$. So one may question the validity of the argument given in [25].}

In the type IIB picture the three-brane BPS mass (per~unit volume) is dilaton
independent (in the Einstein frame) and only depends on the Planck mass (i.e.,
($\sqrt g \alpha^\prime)^{-1/2}$) and three-brane charge.  This is due to the
fact that the three-brane (unlike strings) is a U-duality singlet in IIB
theory.  Therefore, the horizon geometry depends only on the Planck scale and
three-brane charge and not on the asymptotic value of the dilaton.

This is consistent with the attractor mechanism which essentially states that
at the horizon, black brane physics looses memory of the initial values of
moduli fields [30] [41].
\bb

 \no{\bf III. Preliminaries on scalar fields in anti-De Sitter space.}

\b
{\it Coordinates.}

Five-dimensional anti-De Sitter space may be defined as a subspace of ${\bf
R^6}$, more
precisely of a 6-dimensional pseudo Euclidean space with signature
$+,-,-,-,-,+$. Let
$\{y_{i}\}_{i=0,1,...,5}$ be global coordinates of this space, then AdS$_5$
is the sub-manifold
defined by the polynomial equation
$$
y^2 := y_0^2 - \vec y^2 + y_5^2 = \rho^2,~~\vec y := (y_1,...,y_4),\eqno(3.1)
$$
where $\rho > 0$ is a fixed parameter.
It is possible to introduce coordinates, for example $y_0,...,y_4$, but as
any set of coordinates
is of necessity local it is preferable to retain the parameterization in
terms of
$\{y_i\}_{i = 0,...,5}$. Nevertheless, it will be expedient, a little
later, to introduce a
particular time coordinate, $t$.

It will be convenient to generalize: from now on $d=p+1$ will denote the space
dimension,
so that $d = 4$ in the case of present interest. This generalization will
be made only in the
formulas, in the text we always take $d=4$, save explicit emphasis to the
contrary.
\b
\no{\it Boundary asymptotics.}

As a topological space, AdS$_5 = S_3 \times S_1 \times {\bf R^+}$. The
radial coordinate
(the coordinate of ${\bf R}^+$) is defined by
$$
\vec y^2 = r^2,~~ y_0^2 + y_5^2 = \rho^2 + r^2.
$$
The boundary at infinity is the limit $r \rightarrow \infty$, taken for
a fixed point in $S_3 \times S_1$. As the parameter $\rho$ disappears in
the limit, the boundary
can be identified with the cone $y^2 = 0$. More precisely, let $f$ be a
function on AdS; if we
look at $\rho$ as an abreviation for $\sqrt {y^2}$, then $f$ has a unique
extrapolation $F$ to the
subspace $y^2 > 0$ of pseudo-Euclidean 6-space, homogeneous of some
arbitrary but fixed degree,
$N$, say. Then, if the limit exists, we have
$$
\lim_{r\rightarrow \infty}r^{-N}f = (\rho^{-N}F)|_{y^2 = 0}.\eqno(3.2)
$$
The appropriateness of this type of limit will be demonstrated below.

\bb
\no{\it Wave equation.}

The wave equation for a scalar field can be seen as a condition to fix the
value of the second
order Casimir invariant of $so(4,2)$. The vector fields
$$
L_{ij} = y_i\partial_j - y_j\partial_i,
$$
where $\partial_i = \pm \partial/\partial y_i$, the sign reflecting the
signature, act on the
manifold defined in Eq.(1.1) and generate the action on that space of the
group $SO(4,2)$.
The second order Casimir operator is
$$
{\cal C} := -{1\over 2}\sum_{i,j} L_{ij}L_{ij} = - y^2\partial^2 + \hat
N(\hat N+d).\eqno(3.3)
$$ The sum is over all 6 values of the subscripts and $\hat N$ is the
operator associated with the
degree of homogeneity,
$$
\hat N f = y\cdot \partial f = \sum y^i\partial_i f.\eqno(3.4)
$$\b

\no{\it Quantum numbers.}

An elementary particle in five-dimensional anti-De Sitter space  is a unitary,
irreducible, projective, highest weight representation of $SO(4,2)$.
Weights are defined in terms
of finite-dimensional representations of the compact subgroup $SO(4)\otimes
SO(2)$. The generator
of the second factor is $L_{05}$, the {\it energy}. Highest weight means
lowest energy;
the energy is bounded below. Since $SO(4) = SU(2) \otimes SU(2)$, a highest
weight of the first
factor is characterized by a pair $(j_1,j_2)$ of positive half-integers. An
elementary particle
is thus characterized by an irreducible representation $D(E_0;j_1,j_2)$,
where $E_0$ is the lowest
value of the energy. The existence of highest weight representations
implies a discrete energy
spectrum within any unitary, irreducible representation; we normalize the
energy generator,
$$
\hat E = i(y_0\partial_5 - y_5\partial_0),
$$
so that the interval is equal to unity. But the eigenvalues need not be
integers. (Whence the
qualification ``projective" representation.)  There are limitations on
$(E_0,j_1,j_2)$ that
are necessary and sufficient for the representation $D(E_0,j_1,j_2)$ to be
unitary:
$$
E \geq \biggl\{\matrix{ j_1 + j_2 + 2,&j_1j_2 > 0,\cr
j_1 + j_2 +1,& j_1j_2 = 0.\cr} \eqno(2.5)
$$

The highest weight modules will appear as spaces of field modes. A function
associated
with an eigenvalue $E$ of the energy  has the form
$$
f = e^{-iEt}g(\vec y),
$$
where $y_0 - iy_5 = \sqrt{\rho^2 + r^2}e^{it}$ defines the time $t$, and $
\hat E = i{d\over dt}$.
\b
To a highest weight corresponds a value of the Casimir operator, defined in
(3.3),
$$
{\cal C} = E_0(E_0-4) + 2 j_1(j_1 + 1) +2 j_2(j_2+1).\eqno(3.6)
$$
Therefore, the wave equation associated with $D(E_0,0,0)$ is
$$
[y^2\partial^2 - \hat N(\hat N+d) + E_0(E_0-d)]\varphi = 0.\eqno(3.7)
$$
\no{\it Propagator.}

Instead of solving the wave equations for the fields, it will be much more
useful to calculate
the propagators. The reason is that the propagator gives additional
information, concerning the
existence of field quantization. This information is relevant for the
classical field theory as
well, since Greens functions are important for the construction of
solutions in the presence of
external perturbations, and also in connection with questions of
completeness. All this information
can also be culled from the action of the conformal group on the field
modes, but this leads to
a discussion of indecomposable representations, a subject that we prefer to
introduce later.

An invariant propagator is a function of two points labelled $y$ and $y'$
in AdS$_5$, invariant
under the action of SO(4,2). Locally, it is a function $K$, depending on
the variable,
$$
z := y\cdot y' = \sum_i \pm y_iy'_i.
$$
The propagator associated with the wave equation (3.7) satisfies
$$
[(1-z^2)\partial_z^2 - (d+1) z\partial_z + E_0(E_0-d)]K(z) = 0.\eqno(3.8)
$$
The solutions, for most values of the parameter $E_0$, are hypergeometric
series, $K(z)
= z^s\times $ inverse power series. The indicial equation is
$$
(-s+E_0)(s-d+E_0)  = 0,\eqno(3.9)
$$
  Usually, except for special cases to be dealt with,
this signals the presence of two highest weight modules, one with minimal
energy $E_0$ as
expected, and the other with lowest energy $4-E_0$. If $E_0> 3$, then only
the first is unitary,
and since this is the larger of the two, there is a propagator that is  a
hypergeometric series.
Interpreted as a Fourier series, it is the reproducing kernel for an
irreducible representation
of type $D(E_0,0,0)$, and all is well. If $2 < E_0< 3$  nothing untoward
happens, except that now there are two hypergeometric series that solve the
propagator
equation and one may adopt either solution. (But not both!) The domain
$1<E_0<2$ is of course similar.

There remains the most interesting cases. If $E_0 = 2$, then  only one
solution is a hypergeometric
series, and the other solution is logarithmic. This means that the energy
operator cannot be
diagonalized, and signals the appearance of a nondecomposable
representation. This is not a
difficulty, since one simply rejects the logarithmic solution in favor of
the other.

Finally, the case $E_0 = 1$ is also logarithmic, since $s_1-s_2 = 2$. Only
one of the two solutions
is a power series, and it is the wrong one, for it propagates the set of
modes that correspond to
$D(3,0,0)$, with minimal energy $E_0 = 3$.

In conclusion, the case of a scalar field carrying the representation with
minimal energy $E_0 =
1$ is anomalous, since its propagator is logarithmic, namely:
$$
K(z) = z^{-1} + az^{-3} + ... - {\rm log}z (z^{-1} + bz^{-3} + ...).
$$
The wave modes are of three types: physical modes that fall off as $r^{-1}$
at infinity,
``gauge modes" that fall off as $r^{-3}$, and ``scalar modes" that go like
either
$r^{-3}$ or $r^{-3}$log$\,r$. Taken together, they form a nondecomposable
$SO(4,2)$ module. (In a notation to be
explained later, it is of the type $D(3,0,0) \to D(1,0,0) \to D(3,0,0)$.)
The gauge modes form an
invariant submodule with the non-logarithmic propagator and minimal energy
$E_0 = 3$. The scalar modes are conjugate to the gauge modes and appear
paired with them in the Fourier
expansion of the propagator. (Not quite a Fourier series because of the
logarithm, but an
expansion in terms of generalized eigenfunctions of the energy operator.)

The gauge modes have most of the properties of an ordinary massive or
massless fields in 5
dimensions. They form a complete set in the usual sense, which means that
there are enough modes
to make a local measurement; that is, the field can be used to find a small
object. The set of
physical modes, which are physical just because we are trying to use the
representation with
$E_0 = 1$ to do physics, is very different. It is very singular, mainly in
the sense that there is
very little degeneracy in the energy spectrum. In fact, the degeneracy  of
each energy level is
precisely that of an ordinary particle in four space time dimensions and
the black body radiation
has the energy density $\propto T^4$ that is normal for a
three-dimensional space.

This is of course what one should expect. We are looking for a field that
extrapolates from the
boundary to the interior, such that, on the four-dimensional boundary it
describes a massless
particle. In fact, the representation $D(1,0,0)$ of the conformal group of
3-dimensional
Minkowski space is realized precisely as the space of states of an ordinary
massless particle.
What distinguishes this representation of the conformal group is that it
remains irreducible when
restricted to the Poincar\'e subgroup [31].  Here, it may be argued, we are
more concerned with its
reduction to $SO(3,2)$, in as much as the boundary is AdS$_4$ rather than
Minkowski. But the
situation is essentially the same, the restriction of the representation
$D(1,0,0)$ to $SO(3,2)$ is
the  direct sum of just two irreducible representations [32].

To preserve unitarity in the presence of interactions it is necessary that
the interaction be
gauge invariant; that is, it must be insensitive to gauge modes. But this
is a severe requirement,
since gauge modes are so much more plentiful. The only thing that sets the
physical modes apart
is their slow decrease at infinity; precisely, the limit
$$
\lim_{r \rightarrow \infty} r\varphi(y)
$$
is gauge invariant. So, to be gauge invariant, the interactions can be
sensitive only to the
boundary value of the field; in short, the interactions take place at
infinity, exclusively.

This conclusion applies in general, to all the fields examined in this paper.

There is an alternative to admitting the logarithmic modes; it consists of
replacing the second order wave operator
by its square [33]. This admits a propagator of the form of a
hypergeometric series. The non-decomposable
representation, the boundary conditions and the attendant gauge structure
are exactly the same. The difference is
that in this formulation it is clear what the Lorentz (physicality)
condition is: it is just the second order
wave equation.  In the logarithmic formulation that we have chosen in this
paper, in order to remain within the
context of ordinary, second order wave equations, it is not clear whether
the Lorentz condition has a local
expression.

Similar considerations applied to AdS$_{p+2}$ lead to the conclusion that
the most interesting
scalar field carries a representation with minimal energy $E_0 = {p-1\over
2}$. In anology with the analysis of [21], we
expect that such singleton modes should appear in the Kaluza-Klein version
of IIB supergravity compactified on AdS$_5
\times S_5$.

\bb

\no {\bf IV. Conformal Maxwell theory.}

All free fields, of any local field theory, satisfy after gauge fixing the
same wave equation.
In our case, whether the field $A$ with components $A_i$ is a scalar, a
spinor or a vector field,
it satisfies after gauge fixing the equation
$$
[y^2\partial^2 - \hat N(\hat N+d) + E_0(E_0-d)]A_i = 0,~i = 1,..,n~.\eqno(4.1)
$$
The field takes values in an $n$-dimensional $SO(4,2)$ module. Therefore,
it transforms as a
subrepresentation of $D_{{\rm orb}}\otimes D_{{\rm spin}}$, where the
second factor is
$n$-dimensional and the first factor is of the type, containing
$D(E_0,0,0)$, examined in the
preceding section.

Recall that electrodynamics in Minkowski space is carried by a vector field
that satisfies, after
gauge fixing, the wave equation and the Lorentz condition. The space of
solutions of this pair
of equations includes the transverse physical modes and the longitudinal
gauge modes. But this
does not allow for the construction of a quantized field, for  no invariant
propagator can be
constructed from these modes alone.  Nor can we drop the gauge modes since,
though they form an
invariant subspace, there is no invariant complement of physical modes. The
remedy is to drop,
temporarily, the Lorentz condition, and to make use of all the modes that
solve the wave equation;
that is, the entire representation $D_{{\rm orb}}\otimes D_{{\rm spin}}$.

In the last paragraph we had in mind the Poincar\'e group and its
representations. It has been known,
for a very long time, that Maxwell's equations are conformally invariant.
In particular, the
wave equation is invariant, and the solutions carry a representation of
$SO(4,2)$. Unfortunately,
though the wave equation is invariant, the wave operator is not, so to
construct a conformally
invariant propagator one must do something radical. One way that it can be
done is to introduce an
additional scalar field. And the simplest way to get support for this idea
is to pass to
Dirac's formulation of (the conformal completion of) Minkowski space as a
cone in six dimension; this is, as we have seen,
the same as the boundary of our space AdS$_5$.

Dirac's idea [34] was to develop a formalism that makes
conformal invariance manifest; the next step is therefore to replace the
four-dimensional vector
field of ordinary electrodynamics by a six-dimensional vector field. Of
course, all six components
of this field should satisfy (after gauge fixing) the scalar wave equation
(4.1).
Therefore, to start with, we are looking at the representation $D_{{\rm
orb}}\otimes D_6$.
Since we are on the cone, where the Casimir operator does not qualify as a
wave operator,
we try to use the invariant operator $\partial^2$, but something must be
said about the meaning of
this operator.

The problem is that the operators of partial differentiation with respect
to $y_i$ are not
tangential to the cone $y^2 = 0$. Therefore, the operator $\partial^2$ is
not likely to be defined
on the cone. But it turns out that this last operator is nevertheless well
defined if the field is
homogeneous of degree $-1$. This ties in nicely with the considerations of
Section 2.
The equation
$$
\partial^2A_j = \sum_{i = 1,...,5}\pm \partial_i^2 A_j = 0\eqno(4.2)
$$
is meaningful and  conformally invariant. In terms of local coordinates
$\partial^2$ is the
four-dimensional d'Alembertian. The most
disquieting feature is that the field has six components instead of four.
It is therefore to be
expected that subsidiary conditions of the type $y\cdot A = 0$ or $\partial
\cdot A = 0$ must be
imposed. This cannot destroy the  circumstance that the conformal group
acts on the space of
solutions, for each condition projects on an invariant submodule. But it
can destroy the
quantizability of the theory. In fact, quantization requires the retention
of all components;
subsidiary conditions can be imposed only on the states, just as in the
conventional formulation.
This is because the entire representation
$$
D_{{\rm boundary}} = D(1,0,0) \otimes D_6
$$
is nondecomposable; it has no direct {\it summand}. To be precise,
$$
D_{{\rm boundary}} \approx D(1,{\scriptstyle{1\over 2},{1\over 2}})
\rightarrow [D(2,1,0)\oplus D(2,0,1)
\oplus {\rm Id}]
\rightarrow D(1,{\scriptstyle{1\over 2},{1\over 2}}).
$$ The arrows have the following meaning. $A = B\rightarrow C$ means that
the big representation
$A$ has a subrepresentation $C$ with a complement $B$, and that there is no
invariant
complement. ~The physical modes are in the center; note the inclusion of a
zero mode, \break Id $= D(0,0,0)$, the gauge
modes are on the right and the conjugate scalar modes are on the left.
Except for the presence of
the zero mode, and the fact that the spaces of gauge and scalar modes are
larger than usual, this
is exactly the structure of ordinary electrodynamics, as a Poincar\'e
module. (The two representations
in the middle remain irreducible when restricted to the Poincar\'e subgroup.)

Incidentally, the Lorentz (physicality) condition is
$$
y\cdot A = 0. \eqno(4.3)
$$
This analysis of conformal Maxwell theory may be found in [35].

Let us now go to the interior of AdS$_5$, trying the same strategy. That
is, we start with the
same big representation $D(1,0,0) \otimes D_6$ and try to implement it by
imposing the wave
equation (3.1) on each of the six field components. But we have seen that
this does not work;
we do not have a realization of $D(1,0,0)$ in terms of a scalar field. The
only way is to use
the logarithmic propagator for each field component and the propagator
$K_{ij}=\delta_{ij}K_{\log}$
for the vector field.
 This means that each component of $A$ carries the representation of the
scalar singleton field, namely
$$
\tilde D(1,0,0) := D(3,0,0) \to D(1,0,0) \to D(3,0,0),.$$
The vector field thus carries  (to put it differently, the propagator is
the reproducing kernel for)
the monstrous representation
$$\eqalign{
D_{\rm interior} &=  \tilde D(1,0,0) \otimes D_6 \cr
& =[D(3,0,0)\otimes D_6] \rightarrow D_{{\rm boundary}} \rightarrow
[D(3,0,0)\otimes D_6].
\cr}$$
Thus, on top of the usual gauge structure, already somewhat amplified by
the exigency of
conformal invariance, we have a new instance of the  singleton type of
gauge structure,  and
four dimensional Maxwell theory turns into a five dimensional topological
gauge theory.

Let us emphasize this conclusion. Maxwell theory on the boundary of AdS$_5$,
 in ordinary, four dimensional space time, can be extended to the five
dimensional
interior. But the extended field, though it includes extrapolations of all
the physical
modes, are swamped by a much more dense set of new excitations. These extra
modes enter the
propagator with both signs. A Gupta-Bleuler type of quantization is still
possible, but now
there is a new class of gauge modes and a new type of gauge
transformations. There
is no conformally invariant local interaction that preserves gauge
invariance, and therefore there
is no way to introduce interactions, except possibly on the boundary, that
preserves conformal
invariance. In other words, the extrapolated theory is purely topological,
exactly as the
prototype singleton theory in four dimensions [19][20].
\bb
\line{{\bf V. Five-Dimensional Superconformal $U(1)$ Theory.} \hfil}
\vskip.3cm
\def\1/2{{{\scriptstyle{1\over 2}}}}
In this section we will first give a manifest $O(4,2)$ covariant
formulation of supersymmetric $N=1$ and $N=4$ gauge theory in four
dimensions Minkiwski space, conformally extended to the Dirac
cone, and expressed in terms of the natural coordinates of the cone
[24][36]. Then
we discuss the extrapolation to the interior of AdS$_5$.

The $N=4$ theory can be viewed as a $N=1$ theory with three chiral
multiplets added to the $N=1$ gauge multiplet, to build up the $N=4$ vector
multiplet.  In this formulation, the $SU(4), \,N=4$ theory has manifest
$N=1$ supersymmetry and global $SU(3)$ symmetry.  Group-theoretically this
corresponds to the decomposition $SU(2,2/4)\to U(2,2/1)\otimes SU(3)$.
Equipping the field of the chiral multiplet with an index $\alpha (\alpha =
1,2,3)$ we have the
transformation rules [36]:
$$
\eqalign{\delta A^\alpha &= -i\bar\chi\,y\cdot\gamma\,\psi^\alpha,
~~~\hskip1cm
\delta B^\alpha = -i\bar\chi\,y\cdot\gamma\,\psi^\alpha, \cr
\delta F^\alpha &= i\bar\chi(\gamma^{ij}L_{ij}-2)\psi^\alpha, ~~~~~
\delta G^\alpha = i\bar\chi\gamma_7(\gamma^{ij}L_{ij}-2)\psi^\alpha, \cr
\delta\psi^\alpha &= -\gamma\cdot\partial(A^\alpha-\gamma_7B^\alpha)\chi +
(F^\alpha-\gamma_7G^\alpha)\chi, \cr
\delta\lambda &= -\gamma^{ij}G_{ij}\chi - D\gamma_7\chi, \cr
\delta G_{ij} &= i\bar\chi(2\gamma_{ij}+y^k\gamma_{kj}\partial_i
-y^k\gamma_{ji}\partial_j)\lambda, \cr
\delta D &= i\bar\chi(\gamma^{ij}L_{ij}-2)\lambda. \cr}
$$
Here $(\psi^\alpha , \lambda)$ are $O(4,2)$ Majorana 8-spinors, $\{y_i\}$
are the same cone
variables as before, and $\{\gamma_i\}$
are six Dirac matrices. The $L_{ij}$ are the orbital $O(4,2)$ generators.
Finally, $\chi$ is an 8-dimensional, anticommuting
spinor parameter corresponding to Poincar\'e and conformal supersymmetry.

The free-bosonic field equations on the Dirac cone are
$$
\partial^2 A^\alpha = \partial^2B^\alpha = \partial^2A_i = 0
$$
\no where the gauge-potential $A_i~(G_{ij} = \partial_iA_j-\partial_jA_i)$
is subjected to the gauge condition $\partial^iA_i=y^iA_i = 0$.  The
spinor field equations are
$$
(\gamma^{ij}L_{ij}-2)\psi^\alpha = 0 = (\gamma^{ij}L_{ij}-2)\lambda.
$$
\no Note that the chiral $U(1)$ charge of this
$(U(2,2;1)$ algebra is 3/4 for $\lambda$ and -1/4 for the $\psi^\alpha$,
according to [13].  This follows from the fact that  the fields
$A^\alpha, B^\alpha$ and $A_i$ are homogeneous of degree -1, while
$F^\alpha, G^\alpha, \psi^\alpha, D$ and $\lambda$
are homogeneous of degree -2.

Observe that the $U(1)$ charge matrix of the four fermions
$(\psi^i,\lambda)$ is traceless as it must be since
$U(1)\subset SU(4)$. Also the
$U(1)$ charge of the (complex) triplet
$A^i+iB^i$ is ${1\over 2} $as it must be in order that the (non-Abelian)
conformal couplings be $U(1)$ invariant [39].

It is obvious that the above free-field equations (as well as their
non abelian extension to the interacting case) are $SU(4)$ covariant
on the Dirac cone.
In a forthcoming paper the $N=4$ manifest formulation together with
its extension to $AdS_5$ will be given.
It is now possible,
nevertheless, to reach some conclusions concerning the nature of the
extrapolation of super conformal $U(1)$ gauge theory
to the interior of AdS$_5$.

The field equations given above must certainly remain valid in the complete
theory; we mean that the equations $\partial^2
A^\alpha = 0$ and the rest have to be satisfied by the physical modes of
the theory (and by the gauge modes as well).
Since these field equations are to be the limits of field equations
satisfied by the extrapolated fields, we take it to be
evident that this extrapolation will involve exactly the same type of
representations of $SO(4,2)$
that we studied in the earlier sections, though we do not know exactly how
many of them will be carried by the propagator,
assuming that it exists. Therefore, it is quite clear that the physical
exitations, extrapolated from the boundary at infinity to
the interior, are swamped by much more numerous gauge modes, that they are
distinguised from the latter by their boundary
conditions at infinity only, and not by any local property, and that
consequently they cannot be observed. To put it somewhat
differently, the full (``Gupta-Bleuler") quantization space has an
indefinite metric, unitarity therefore requires that the
interactions be gauge invariant in the sense that they do not engage (what
we have called) the gauge modes. And since the
gauge modes cannot be characterized locally, this implies that there cannot
be any local interactions at all.

It is possible to calculate the smallest representation of conformal
supersymmetry ($N = 1$) that contains all the physical
modes and that must be involved in any complete, off-shell formulation of
the theory. To describe the
representation, let
$D^s(E_0,j_1,j_2/c)$ denote the unique, irreducible super symmetry module
with highest weight $(E_0,j_1,j_2/c)$, where the last
label refers to the
$u(1)$ charge. Then the total representation must contain
$$
  D^s(\1/2,\1/2,0/\1/2) \rightarrow [D^s({\scriptstyle{3\over
2}},0,\1/2/\1/2) \oplus {\rm Id}
\oplus D^s({\scriptstyle{3\over 2}},\1/2,0/-\1/2)] \rightarrow
D^s(\1/2,0,\1/2/-\1/2).
$$
As usual, physical modes are in the middle and gauge modes are at left.

We leave the complete construction of the supersingleton field theory
of the $SU(2,2/4)$ algebra to a forthcoming work [40].

 \bb

\line{{\bf VI.  Conclusions.} \hfil}
\vskip.3cm

Motivated by recent attempts to relate world-volume dynamics to supergravity
in the near horizon geometry, we have analyzed in some detail some
features of the ``dual theories" underlying 3-brane dynamics in IIB string
theory.  This amounts to comparing gauged $D=5$ supergravity, with supergroup
$SU(2,2/4)$ and $N=4$, $SU(4)$ super Yang-Mills\break theory.  On the
supergravity side we
find that a proposed quantization condition on the coefficient of the
gauged Chern-Simons coupling
[25] in the horizon geometry is equivalent to the statement that the horizon
geometry of the three-brane is independent from the dilaton and only depends on
the Planck scale, similar to the attractor mechanism [30].

On the world volume side, we have analyzed a superconformal $U(1)$ gauge
theory and have shown that
it can be formulated as a topological (singleton) field theory in
AdS$_5$, in close parallell to what happened with the five-brane in $M$
theory compactification [6][10][21].

For the future, we propose to complete the investigation of  $N = 4$,
superconformal, $U(1)$
gauge theory initiated in Section V, and to attack the problem of the relation
of the interacting $SU(N)$ nonabelian $N = 4$ gauge theory with anti-De
Sitter geometry.
It is at present uncertain whether
brane dynamics (on the world volume) can be related to singleton field
theories on anti-De Sitter boundaries, but we
think that further work in this direction may lead to new and interesting
duality connections.

We hope the considerations  presented in this paper, some of a purely
speculative nature,  will help to shed new
light on these actual and potential connections among different aspects of
non-perturative string dynamics.
\bb

\no{\steptwo Acknowledgements.}

We would like to thank R. Kallosh, R. Stora and A. Zaffaroni for
interesting discussions. S.F. is
supported in part by DOE
under grant DE-FG03-91ER40662, Task C, and by EEC Science Program
SCI$*$-CI92-0789 (INFN-Frascati).

\bb

\line{{\bf References.} \hfil}
\vskip.3cm

\item {[1]} See, e.g., A Salam and E. Sezgin, ``Supergravity in Diverse
Dimensions." (World Scientific) (1989) Vols. I and
II.

\item {[2]} E. Witten, Nucl. Phys. {\bf B443} (1995) 85.

\item{[3]} J. Maldacena, hep-th/9711200.

\item {[4]} M.P. Blencowe and M.J. Duff, Phys. Lett. {\bf B203} (1988) 229;
Nucl. Phys. {\bf B310} (1988) 389;
M.J. Duff, Class. Quant. Gravity, {\bf 5} (1988) 189.
E. Bergshoeff, M.J. Duff, C.N. Pope and E. Sezgin, Phys. Lett. {\bf B199}
(1988) 69.

\item {[5]}H. Nicolai, E Sezgin and Y Tanii, Nucl. Phys {\bf B305} (1988) 483.

\item{[6]} C. W. Gibbons, P. K. Townsend, Phys. Rev. Lett. {\bf 71} (1993)
3754.

\item{[7]} G. Gibbons, Nucl. Phys. B{\bf 204} (1982) 337.

\item{[8]} A. Chamseddine, S. Ferrara, G. Gibbons and R. Kallosh, Phys.
Rev. D{\bf 55} (1997) 364.

\item {[9]} E. Cremmer, B. Julia and J. Scherk, Phys Lett. {\bf B76} (1978)
409.

\item{[10]} P. Claus, R. Kallosh, and A. van Proeyen, hep-th/9711161;
R. Kallosh, J. Kumar, A. Rajaraman, hep-th/9712073.

\item {[11]} W. Nahm, Nucl. Phys. {\bf B135} (1978) 149.

\item {[12]} R. Haag, J. Lopuszanski and M. Sohnius, Nucl. Phys. {\bf B88}
(1975) 257.

\item {[13]} J. Wess and B. Zumino, Nucl. Phys. {\bf B77} (1974) 73.

\item{[14]} S. Ferrara, M. Kaku, P. Townsend, P. van Nieuwenhuizen.
Nucl. Phys. B{\bf 129} (1977) 125.

\item{[15]} M. Douglas, J. Polchinsky, A. Strominger, hep-th/9703031.

\item{[16]} E. Bergshoeff, A. Salam, E Sezgin and Y. Tanii, Nucl. Phys.
{\bf B305} (1988) 497.

\item {[17]} M. Gunaydin, B.E.W. Nilsson, G. Sierra and P. Townsend, Phys
Lett. {\bf 176B} (1986) 45.

\item {[18]} E. Bergshoeff, A. Salam, E. Sezgin and Y. Tanii, Phys. Lett.
205 (1988) 237.

\item{[19]}   M. Flato and C. Fronsdal, J. Math. Phys. {\bf 22} (1981) 1100.

\item{[20]} H. Nicolai and E Sezgin, Phys. Lett. {\bf 143B} (1984) 389.

\item{[21]} M. Gunaydin, P. van Nieuwenhuizen, N. P. Warner, Nucl. Phys.
B{\bf 255} (1985) 63.

\item {[22]} B. de Wit and H. Nicolai, Nucl. Phys. {\bf B208} (1982) 323.

\item{[23]} M. Pernici, K. Pilch, P. van Nieuwenhuizen, Phys. Lett.
B{\bf 143} (1984) 103.

\item{[24]} G. Mack, A. Salam, Am. Phys. {\bf 53} (1969) 174.

\item{[25]} M. Gunaydin, L. J. Romans, N. P. Warner, Phys. Lett. {\bf 154B}
(1985) 268.
See also M. Pernici, K. Pilch, P. van Nieuwenhuizen, Nucl. Phys. {\bf B259}
(1985) 460

\item{[26]} L. Andrianopoli, R. D'Auria, S. Ferrara, P. Fr\'e, R. Minasian,
and M. Trigiante, hep-th/9612202.

\item {[27]} See e.g. E. Cremmer ``Supergravity 81". S. Ferrarra and J.G.
Taylor, page 313; B. Julia in
``Superspace and Supergravity"
edited by S. N. Hawking and M. Rocek, Cambridge (1981) page 331.

\item {[28]} M. B. Green and J. H. Schwarz, Phys Lett. {\bf B122} (1983)
143; J. H. Schwarz, Nucl. Phys. {\bf B266}
(1983) 269; P. S. Howe and J. S. Schwarz, Nucl. Phys. {\bf B
23} (1984) 181.

\item {[29]} P. Howe, K. Stelle and P. Townsend, Nucl. Phys. {\bf B192}
(1981) 332.

\item{[30]} S. Ferrara, R. Kallosh, A. Strominger, Phys. Rev. D{\bf 52}
(1995) 5412; A. Strominger, Phys. Lett. B{\bf 383} (1996) 39;
S. Ferrara, R. Kallosh, Phys. Rev. D{\bf 54} (1996) 1514.

\item {[31]} G. Mack and I. T. Todorov, J. Math. Phys. {\bf 10}(1969) 2078.

\item {[32]} E. Angelopoulos, M. Flato, C. Fronsdal and D. Sternheimer,
Phys. Rev. {\bf 23} (1981) 1278.

\item {[33]}M. Flato and C. Fronsdal, Commun. Math. Phys. {\bf 108} (1987) 469.

\item {[34]} P. A. M. Dirac, Ann. Math. {\bf 37} (1936) 429.

\item {[35]} B. Binegar, C. Fronsdal and W. Heidenreich, J. Math. Phys.
{\bf 24} (1983) 2828.

\item{[36]} S. Ferrara, Nucl. Phys. B{\bf 77} (1974) 73.

\item {[37]} C. Fronsdal, Phys. Rev. D {\bf 26} (1982) 1988.

\item {[38]} M. Gunaydin and N. Marcus, Class. Quant. Grav. {\bf 2} (1985) L11.

\item {[39]} L. Brink, J. H. Schwarz and J. Scherk, Nucl. Phys. {\bf B121}
(1977) 77;
 F. Gliozzi, J. Scherk and D. Olive, Nucl. Phys. {\bf B122} (1978) 1253.

\item {[40]} S Ferrara and C. Fronsdal, in preparation.

\item {[41]} I. Klebanov, Nucl. Phys. {\bf B496} (1997) 231; S. Gubser, I
Klebanov and A. Tseytlin, Nucl. Phys. {\bf B499} (1997)
41.

\item {[42]} K. Sfetsos and K. Skenderis, hep-th/9711138;
S. Hyun, hep-th/9704005;
 H. Boonstra, B. Peeters and K. Skenderis, hep-th/9801076,
 {\it Phys. Lett.} {\bf B411} (1997) 59.

\item{43]} H. J. Kim, L. J. Romans and P. van Nieuwenhuizen, 
``The Mass Spectrum of Chiral $N=2$ $D=10$ Supergravity on $S^5$'',
 Phys. Rev. {\bf D32} (1985) 389.

\end